\documentclass[a4paper]{jpconf}
\usepackage{amsmath}
\usepackage{amssymb}
\usepackage{hyperref}
\usepackage{bm}
\usepackage{tikz}
\usetikzlibrary{shapes.callouts,decorations.pathmorphing}
\usetikzlibrary{arrows,shapes}
\definecolor{uibred}{RGB}{167, 38, 47}
\definecolor{MyDarkGreen}{rgb}{0.0, 0.30, 0.00}
\definecolor{MyDarkBlue}{rgb}{0.0, 0.00, 0.7}
\def\ubr#1#2{\underbrace{#1}_{\text{#2}}}

\def\x{\bm{x}}
\def\k{\bm{k}}
\def\p{{\bm{p}}}
\def\k{\bm{k}}
\def\ubr#1#2{\underbrace{#1}_{\text{#2}}}
\newcommand{\Fig}[1]{Fig.~\ref{#1}}
\newcommand{\Eq}[1]{Eq.~\ref{#1}}
\begin{document}
\title{Initial conditions for hydrodynamics from weakly coupled pre-equilibrium 
evolution}

\author{Aleksas Mazeliauskas}
\ead{aleksas.mazeliauskas@stonybrook.edu}
\address{Department of Physics and Astronomy, Stony Brook University, Stony 
Brook, New York 11794, USA}
\begin{abstract}
We use leading order effective kinetic theory
to simulate 
the pre\nobreakdash-equilibrium evolution of transverse energy and flow perturbations in heavy-ion collisions. We provide 
a Green function which propagates the initial perturbations of the 
energy-momentum tensor to a time
when hydrodynamics becomes applicable. With this map,
the pre-thermal evolution from saturated nuclei to hydrodynamics can be modeled in the framework of weakly coupled QCD.
\end{abstract}

\section{Introduction}

Viscous relativistic hydrodynamic simulations of heavy ion collisions at the BNL
 Relativistic Heavy Ion Collider and the CERN Large Hadron Collider have shown 
 tremendous success in describing simultaneously many of the soft hadronic 
 observables, 
 however initial conditions for hydrodynamics remain one of the largest 
 uncertainties in hydrodynamic modeling of heavy ion collisions 
 \cite{Heinz:2013th,Luzum:2013yya,Teaney:2009qa,Antinori:2016zxe}. In this work 
 I use the  effective kinetic theory of weakly coupled quasi-particles to study 
 the equilibration and the onset of hydrodynamics in heavy ion collisions 
 ~\cite{Kurkela:2015qoa,Keegan:2016cpi}. 
In particular, I focus on the transverse perturbations, which are expected to initiate flow during the equilibration process~\cite{Vredevoogd:2008id}.

\section{Separation of scales}
\label{scales}
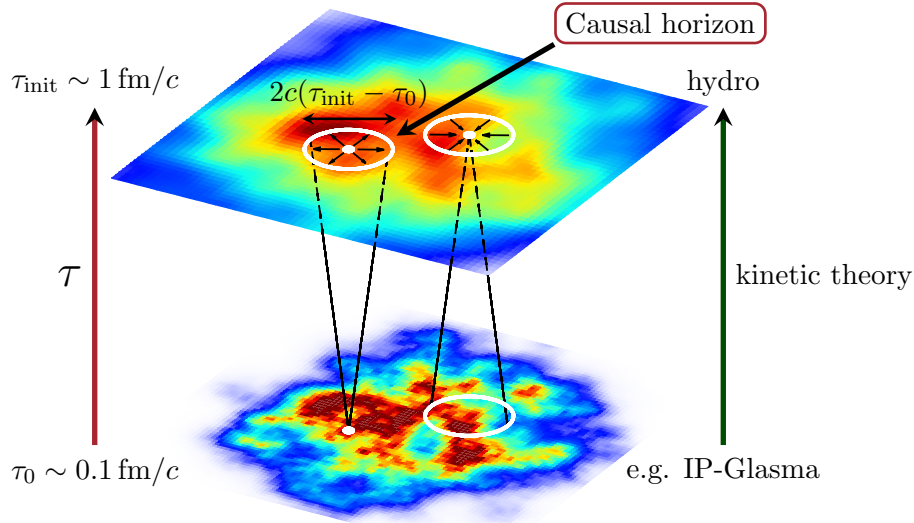
\begin{figure}
\centering
\begin{tikzpicture}[>=stealth]
{\node[anchor=south west] (image) at (0,0) 
{\includegraphics[width=0.5\linewidth]{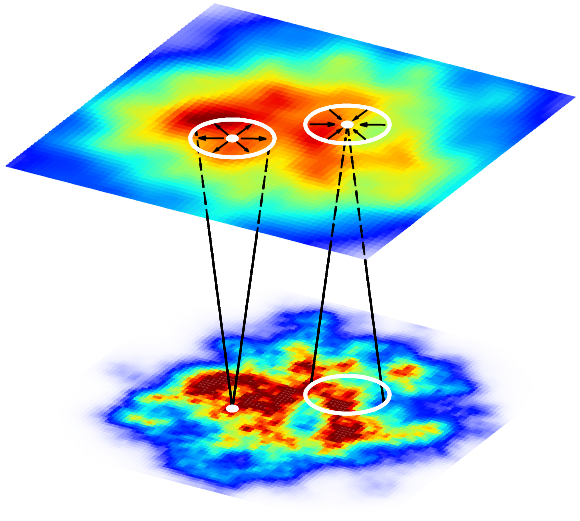}};}
\begin{scope}[x={(image.south east)},y={(image.north west)}]
\draw[->,line width = 2pt, draw=uibred, overlay]  (0,0.2) node[below] {$\tau_0 
\sim 0.1\, \text{fm}/c$} to node[left] {\Large $\tau$} (0,0.8) node[above] 
{$\tau_\text{init}\sim 1\,\text{fm}/c$};
{\draw[->,line width = 2pt, draw=MyDarkGreen,overlay]  (1,0.2) 
node[below] {e.g. IP-Glasma} to node[right  ] { 
kinetic theory} (1,0.8) node[above] { 
hydro};}
{
\draw[<->,line width = 1pt,draw=black] (0.33,0.78) to  node[above] { $2c(\tau_\text{init}-\tau_0)$} (0.48,0.7 8);
\node[overlay, draw=uibred, rectangle, rounded corners, line width=1pt] (c) at 
(0.9,0.95) {Causal horizon}; 
\draw[->, line width = 2pt] (c.south west) to (0.48,0.75);
}
\end{scope}
\end{tikzpicture}
\caption{Kinetic theory describes the evolution from the microscopic formation 
time  $\tau_0$ to the equilibration time $\tau_\text{init}$, when hydrodynamics 
becomes applicable~\cite{Kurkela:2016vts}. By causality, for a given point in 
the transverse plane it is sufficient to analyze the pre-equilibrium evolution 
within the causal neighborhood of that point.\label{causal}}
\end{figure}

In the weak coupling limit,  kinetic theory describes the  evolution of the 
system 
from the microscopic formation time $\tau_0\sim Q_s^{-1}$ to the onset of 
hydrodynamics at a much later time $\tau_\text{init}\sim 
\tau_\text{equilibrium}$~\cite{Kurkela:2016vts}. For realistic values of 
coupling constant $\alpha_s\sim0.3$, the equilibration time is short and the 
causally connected region $c(\tau_\text{init}-\tau_0)\sim 1\,\text{fm}$   is 
much smaller than 
the transverse nuclear geometry 
$R_\text{Pb}\sim 5\,\text{fm}$, but comparable to a single nucleon scale 
$R_p\sim 1\,\text{fm}$ (see 
\Fig{causal} )
\begin{equation}
R_\text{Pb}\gg c(\tau_\text{init}-\tau_0) \sim R_\text{p}. 
\end{equation}
For this reason,
the global nuclear geometry contributes a small gradient to a locally constant 
background, while event-by-event nucleon fluctuations are suppressed by 
$1/\sqrt{N_\text{part}}$, where $N_\text{part}$ is the number of participant 
nucleons. Therefore initial energy density can be expanded locally as
\begin{equation}
e(\x,\tau_0) = e(\tau_0) +\delta e(\x,\tau_0),\label{eq:region}
\end{equation}
where $e(\tau_0)=\left< e(\x,\tau_0)\right>_{|\x-\x_0|\leq 
c(\tau_\text{init}-\tau_0)}$ is the average energy density in the causal region 
and $\delta e(\x, \tau_0)$ is a small perturbation.
We will use effective kinetic theory described below to simulate a 
translationally invariant background with linearized perturbations in the 
causal region $|\x-\x_0|<c(\tau_\text{init}-\tau_0)$.

\section{Effective kinetic theory}

We use the effective kinetic theory of high temperature QCD at leading order in 
$\alpha_s$ to model the pre-equilibrium evolution in heavy ion 
collisions~\cite{Arnold:2002zm}. At early times gluons dominate over fermions 
in the plasma,
so we solve the Boltzmann equation for boost invariant gluon distribution 
function $f(\tau,\x,\p)$ with leading order elastic $2\leftrightarrow2$ and 
inelastic $1\leftrightarrow2$ collision 
processes~\cite{Kurkela:2015qoa,Keegan:2016cpi,Arnold:2002zm}
\begin{align}
\partial_\tau f +\frac{\p}{|p|}\cdot\nabla f- 
\ubr{\frac{p_z}{\tau}\partial_{p_z} f}{\normalsize Bjorken expansion}= 
-\underbrace{\mathcal{C}_{2\leftrightarrow2}[f]}_{\includegraphics[height=0.08\linewidth]{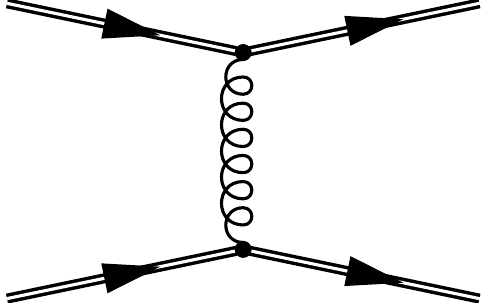}}-\underbrace{\mathcal{C}_{1\leftrightarrow2}[f]}_{\includegraphics[height=0.08\linewidth]{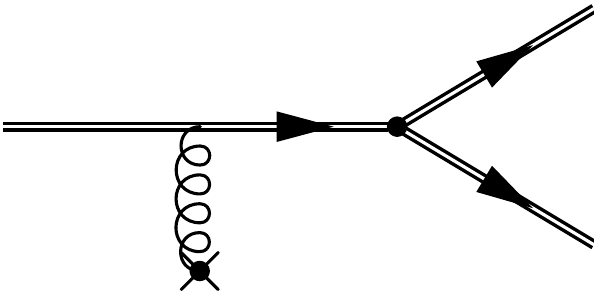}}.
\end{align}
We split the  distribution function into translationally invariant background 
$\bar{f}(\tau,\p)$ and  a linear perturbation with a wavenumber $\k_\perp$ in 
the transverse plane $\delta f_{\k_\perp}(\tau,\p)e^{i\k_\perp\cdot\x}$. Then we
solve the Boltzmann equation as a system of coupled differential equations with 
constant $\k_\perp$.

\begin{figure}
\centering
\begin{tikzpicture}[>=stealth]
\node[anchor=south west] (image) (0,0) {\includegraphics[ 
width=0.45\linewidth]{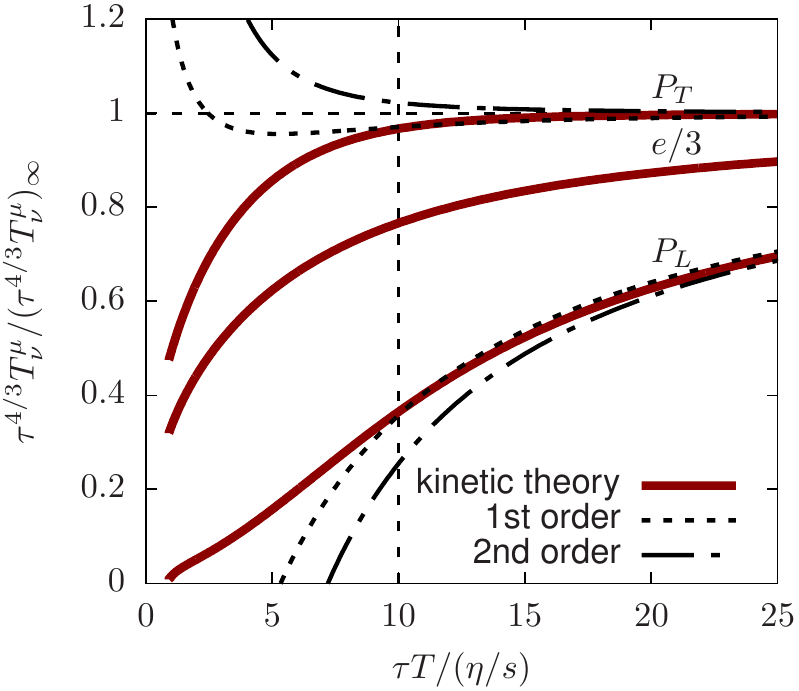}
\includegraphics[width=0.45\linewidth]{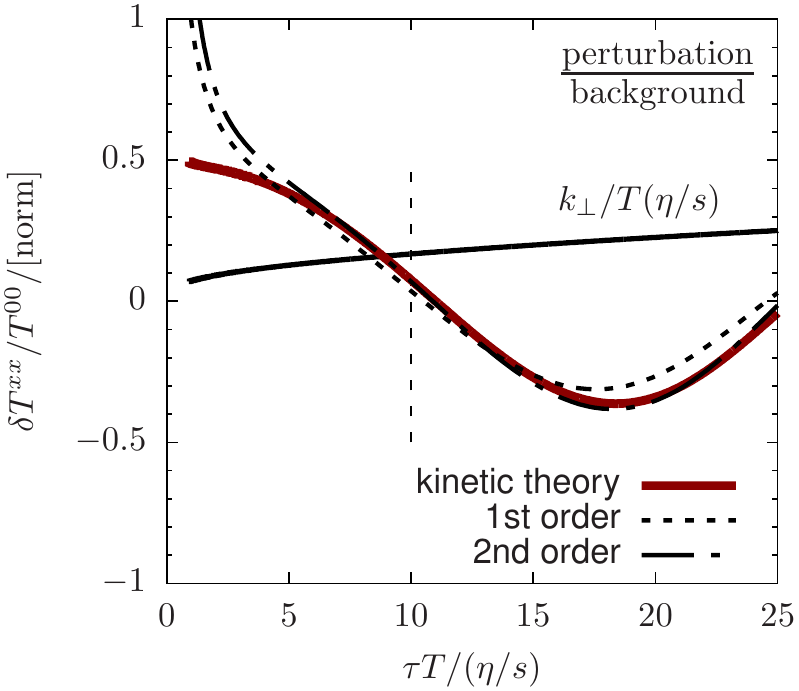}
};
\begin{scope}[x={(image.south east)},y={(image.north west)}]
{\draw[line width=3pt, draw=MyDarkBlue, ->] (0.45,0.8) to node[above, 
fill=white] {hydro} (0.86,0.8)  ;
\draw[line width=3pt, draw=MyDarkGreen, ->] (0.18,0.8) to node[above, 
fill=white] {kinetic\phantom{y}} (0.45,0.8);}
\end{scope}
\end{tikzpicture}%
\caption{(a) Equilibration of background energy momentum tensor in kinetic 
theory relative to the asymptotic values. Dashed lines correspond to the 
asymptotic first and second order hydrodynamic constitutive equations. (b) 
Evolution of energy momentum tensor perturbation $\delta T^{xx}$ relative to 
the background energy density $T^{00}$ due to wavenumber $k_\perp$ energy 
perturbation in the transverse plane (in $x$-direction).   \label{fig:Tmunu} }
\end{figure}

We use a  Color Glass Condensate (CGC) inspired initial background distribution 
function $\bar{f}(\p)$, which possesses large initial pressure anisotropy, 
$P_T\ll P_L$, and  take the functional form of transverse perturbations to be 
$|\delta f_{\k_\perp}| \sim p\partial_p 
\bar{f}(\p)$~\cite{Kurkela:2015qoa,Keegan:2016cpi}. In \Fig{fig:Tmunu}(a) we 
show the evolution of the background energy momentum tensor components relative 
to their asymptotic values in scaled time $\tau T/(\eta/s)$.
We see that at sufficiently late times the longitudinal pressure $P_L$ in 
kinetic theory 
approaches the constitutive equation of viscous conformal hydrodynamics for 
Bjorken expansion~\cite{Baier:2007ix}
\def\etas{{\tfrac{\eta}{s}}}
\def\scaledtime{\left(\frac{{\tfrac{\eta}{s}}}{\tau T}\right)}
\def\lambdataupi{{\tfrac{\lambda_1}{(\tau_\pi \eta)}}}
\def\taupietasT{{\tfrac{\tau_\pi}{{\eta}/{(sT)}}}}
\begin{equation}
\frac{P_L}{e} = \ubr{\frac{1}{3}}{ideal}- 
\ubr{\frac{16}{9}\frac{\eta/s}{\tau T}}{1st order}-\ubr{\frac{32}{27} 
\frac{\tau_\pi}{\eta/(sT)}\left(1-\frac{\lambda_1}{\tau_\pi\eta}\right) 
\left(\frac{\eta/s}{\tau T}\right)^2}{2nd order} + \ldots\label{eq:const}
\end{equation}
Therefore at late times the system evolution can be smoothly passed from 
kinetic theory 
to 
 hydrodynamics.
In \Fig{fig:Tmunu}(b) we show that the linear perturbations of the energy 
momentum tensor $\delta T^{\mu\nu}$ can   also be described with hydrodynamic 
constitutive equations similar to \Eq{eq:const}  at sufficiently late times and 
sufficiently small wavenumbers $k_\perp$.

\section{Linear response functions}

The goal of the pre-equilibrium evolution is to construct the initialization conditions for hydrodynamics at $\tau_\text{init}$ from a given initial state at $\tau_0$.  Close to equilibrium the full energy momentum tensor $T^{\mu\nu}$ can be constructed via hydrodynamic constitutive equations from the local energy $e+\delta e$ and momentum $\vec{g}$ densities, therefore we only need to know energy and momentum response functions to initial conditions.
\begin{figure}
\centering
\begin{tikzpicture}[>=stealth]
\node[anchor=south west] (image) at (0,0) 
{ 
{\includegraphics[width=0.49\linewidth]{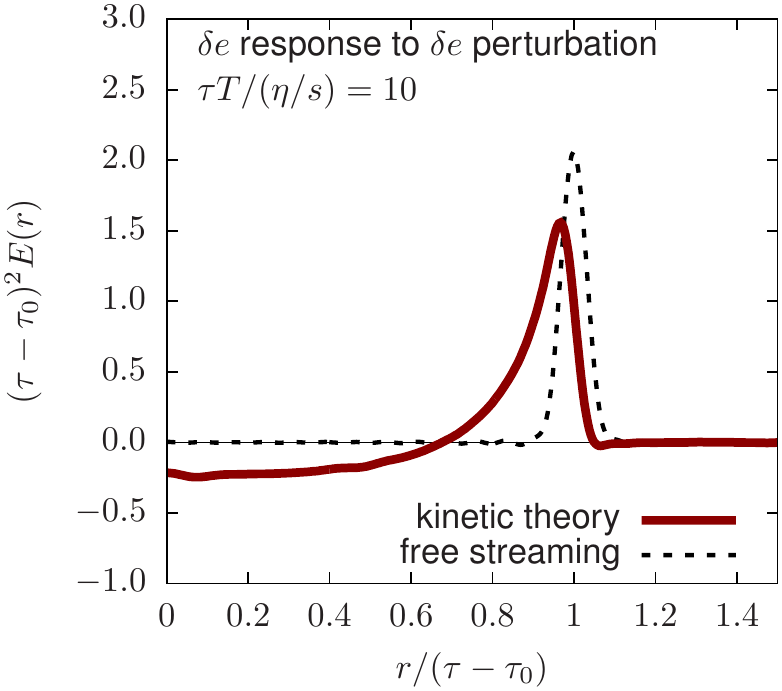}}};
\node[xshift=1cm, yshift=0.cm, align=center, anchor=west] (b) at (image.east) 
{$E(\x;\tau,\tau_0)$\\\includegraphics[width=0.45\linewidth]{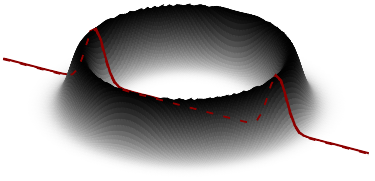}};
\end{tikzpicture}
\caption{(a) Radial profile of energy response function $E(r; \tau,\tau_0)$ due 
to initial energy perturbation at time $\tau T/(\eta/s)=10$. (b) Illustration 
of coordinate space response function $E(\x;\tau,\tau_0)$  to initial 
$\delta$-like perturbation in two dimensions.\label{fig:Green}}. 
\end{figure}

In kinetic theory the linear energy response to the initial energy perturbations can be written as a convolution
\begin{equation}
\frac{\delta e(\tau_\text{init},\x_0)}{e(\tau_\text{init})} = 
\int\! d^2\x \,\,E( |\x_0-\x|;\tau_\text{init},\tau_0)\times \frac{\delta 
e(\tau_0,\x)}{ e(\tau_0) },
\end{equation}
where $E(r;\tau_\text{init},\tau_0)$  is the coordinate space Green function for energy perturbations. In \Fig{fig:Green}(a), we show the radial profile of 
 $E(r;\tau,\tau_0)$  at $\tau T/(\eta/s)=10$ and compare it with the free 
 streaming response. The coordinate space Green function has the meaning of a 
 system response to initial $\delta$-like perturbation (see 
 \Fig{fig:Green}(b)). In the absence of collisions, disturbances propagate at 
 the speed of light and the free streaming response function shown in 
 \Fig{fig:Green}(a) is centered on the causality circle $r=c(\tau-\tau_0)$. 
 As seen in \Fig{fig:Green}(a) collision processes modify the system response 
 in kinetic theory and 
 eventually it will become identical to the hydrodynamic response (not shown).

Spatial Green functions are obtained by simulating the kinetic theory response 
to 
several values of wavenumber $\k_\perp$ perturbations and then taking 
Fourier\nobreakdash-Hankel transform to the coordinate space. Similarly, we 
find 
momentum response function to the initial energy gradients in the transverse 
plane~\cite{Keegan:2016cpi}.

\section{Summary}
We used effective kinetic theory to study  equilibration and approach to 
hydrodynamics of linearized transverse energy perturbations around an initially 
anisotropic but boost invariant background. At the hydrodynamic initialization 
time all components of energy momentum tensor can be initialized from the local 
energy and momentum densities, which can be determined from the kinetic theory 
response to initial perturbations.
Using the kinetic theory pre-equilibrium evolution in heavy ion collision 
models could reduce the dependence on hydro initialization time and better 
account for the pre-equilibrium flow.

\ack
I gratefully acknowledge my collaborators on this work: Liam Keegan, Aleksi 
Kurkela and Derek Teaney. I would like to thank Jean-Fran\c cois Paquet  and 
S\"oren Schlichting for useful comments. Special thanks to the APS Forum on 
Graduate Student Affairs for supporting my participation in the Hot Quarks 2016 
conference. 
Finally, I would like to thank the organizers and participants of the  Hot 
Quarks 2016 conference for the stimulating week of talks and discussions.
This work was supported in part by the U.S. Department of Energy under 
Contracts No. DE\nobreakdash-FG\nobreakdash-88ER40388.

\section*{References}
\bibliography{preflow}

\end{document}